\documentclass[conference]{IEEEtran}
\IEEEoverridecommandlockouts
\usepackage{cite}
\usepackage{amsmath,amssymb,amsfonts}
\usepackage{algorithmic}
\usepackage{graphicx}
\usepackage{textcomp}
\usepackage{multirow}

\usepackage{makecell}
\usepackage{pifont}
\usepackage{xcolor}
\usepackage{balance}

\def\BibTeX{{\rm B\kern-.05em{\sc i\kern-.025em b}\kern-.08em
    T\kern-.1667em\lower.7ex\hbox{E}\kern-.125emX}}
\begin{document}

\title{Edge Storage Management Recipe with Zero-Shot Data Compression for Road Anomaly Detection
\thanks{This work has been submitted to the IEEE for possible publication. Copyright may be transferred without notice, after which this version may no longer be accessible.}
}

\author{\IEEEauthorblockN{YeongHyeon Park}
\IEEEauthorblockA{\textit{SK Planet Co., Ltd.} \\
Seongnam, Rep. of Korea \\
yeonghyeon@sk.com}
\and
\IEEEauthorblockN{Uju Gim}
\IEEEauthorblockA{\textit{SK Planet Co., Ltd.} \\
Seongnam, Rep. of Korea \\
gim.uju1217@sk.com}
\and
\IEEEauthorblockN{Myung Jin Kim}
\IEEEauthorblockA{\textit{SK Planet Co., Ltd.} \\
Seongnam, Rep. of Korea \\
myungjin@sk.com}
}

\maketitle

\begin{abstract}
Recent studies show edge computing-based road anomaly detection systems which may also conduct data collection simultaneously.
However, the edge computers will have small data storage but we need to store the collected audio samples for a long time in order to update existing models or develop a novel method.
Therefore, we should consider an approach for efficient storage management methods while preserving high-fidelity audio.
A hardware-perspective approach, such as using a low-resolution microphone, is an intuitive way to reduce file size but is not recommended because it fundamentally cuts off high-frequency components.
On the other hand, a computational file compression approach that encodes collected high-resolution audio into a compact code should be recommended because it also provides a corresponding decoding method.
Motivated by this, we propose a way of simple yet effective pre-trained autoencoder-based data compression method.
The pre-trained autoencoder is trained for the purpose of audio super-resolution so it can be utilized to encode or decode any arbitrary sampling rate.
Moreover, it will reduce the communication cost for data transmission from the edge to the central server.
Via the comparative experiments, we confirm that the zero-shot audio compression and decompression highly preserve anomaly detection performance while enhancing storage and transmission efficiency.
\end{abstract}


\section{Introduction}
\label{sec:intro}
\begin{figure}[t]
    \resizebox{\columnwidth}{!}{%
        \hspace*{-0.45cm}
        \begin{tabular}{cc}
            \includegraphics*[width=0.452\columnwidth,trim={0.0cm 2.0cm 0.0cm 0.5cm},clip]{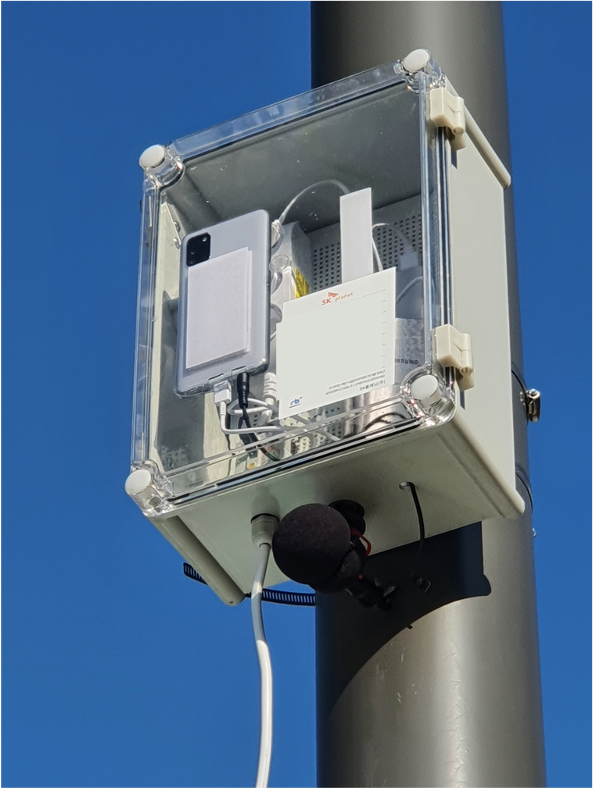} 
            & \includegraphics*[width=0.548\columnwidth,trim={0.0cm 1.2cm 0.0cm 0.0cm},clip]{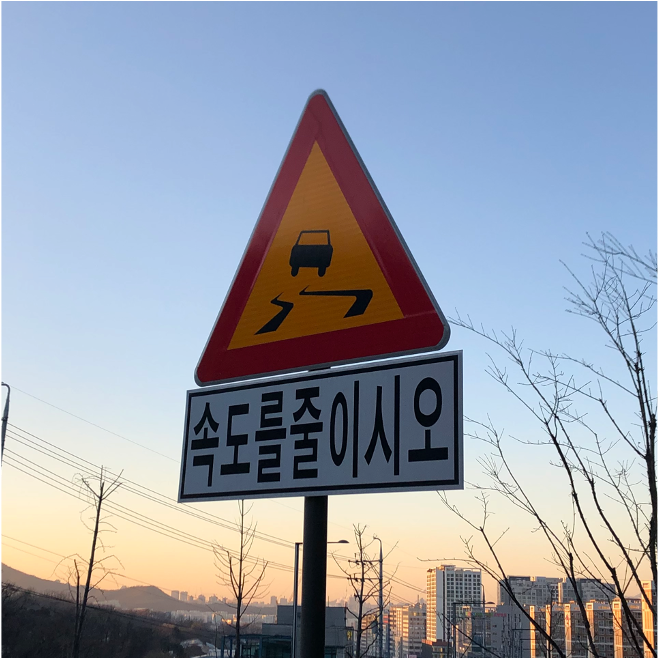} \\
            (a) Edge computer & (b) Road sign for hazard warning \\
            \\
            \multicolumn{2}{c}{\includegraphics*[width=1.0\columnwidth,trim={0.0cm 0.0cm 0.0cm 0.0cm},clip]{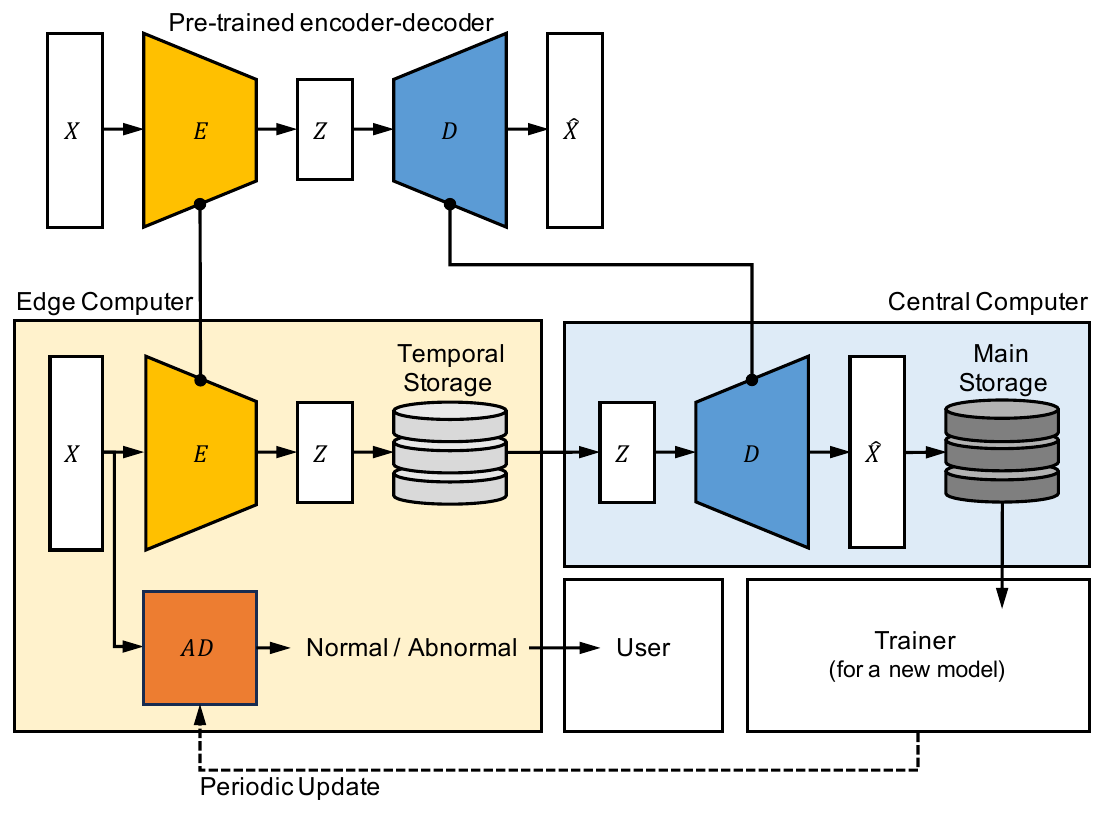}} \\ 
            \multicolumn{2}{c}{(c) Pre-trained encoder-decoder-based storage management system} \\
        \end{tabular} \\
        \hspace*{-0.45cm}
    }
    \caption{Proposed scheme of edge storage management system for road anomaly detection. An edge computer (a) that has a small storage space is installed on the habitual freezing section (b). A proposed method for saving as much data as possible in the limited storage space is shown in (c). Among the pre-trained encoder-decoder, the encoder and decoder are deployed on the edge and central computer respectively. Encoder on edge computer encodes the collected high-resolution audio into latent code Z and the saved Zs are transmitted to the central server at regular intervals. The central server decodes the received latent codes for original high-resolution audio.}
    \label{fig:scheme}
\end{figure}

Knowing road conditions is an effective way to prevent traffic accidents~\cite{park_ncae_2022}. 
Most of the road hazards are highly related to icy or wet roads which reduce the friction between the road and tires.
When considering a vision sensor-based road anomaly detection system~\cite{ryu_camera_2015,rui_camera_2019, bibi_camera_2021,vojir_segmentation_2021}, inclement weather will make occlusion on the camera which makes it difficult for understanding road conditions~\cite{ershadi_camcond_2017,qian_lidar_2021}.
Moreover, intensity-changing situations such as at night also make it difficult to determine road conditions.

As an approach to solving these problems, an audio-based anomaly detection approach has been developed. 
The audio-based system receives information from the medium wave in the air, so it shows a better response-ability than the occlusion situation of the vision sensor.
In addition, sound can be properly transmitted even at night time, an audio-based system is recommended for this situation.

Even in the case of successful anomaly detection as above, continuous updating of the anomaly detection model is required considering that target environments are continuously aging~\cite{song_aging_2023}. 
For this, we need high-quality large data to update the neural network-based anomaly detection model. 
We have installed edge computers on the road for anomaly detection, which has limited resources such as storage.
Considering the above, we have a limitation to keep audio data for the long term.
The above limitation can be partially mitigated by transmitting the audio to large central storage in time, but in this case, enormous data transmission costs will be incurred~\cite{azar_tradcomp_2019}.
Also, we can lower the quality of the audio collected from high fidelity (Hi-Fi) to low fidelity (Lo-Fi) to reduce the file size for each audio, but this is not recommended as it leads to fundamental information loss.

Motivated by this, we propose a storage management method that can keep as much data as possible in the edge computer for a long time in limited storage space while minimizing the cost of the data transmission into a central server. 
Our method is based on zero-shot encoding and decoding using a pre-trained audio super-resolution (ASR) model~\cite{li_AST_unet_2022,han_NuWave_2022,alex_EnCodec_2022}.
Referring ASR models can convert input audio to high-resolution, so they can handle arbitrary resolution inputs that are lower than the target resolution they are trained on.

The overall scheme of our proposal is shown in Fig~\ref{fig:scheme}.
The edge computer continuously collects data and determines whether the situation is abnormal or not through a microphone facing the road.
Basically, collected audio is stored in the original resolution, but it is saved after being converted into a latent vector by an encoder of pre-trained ASR.
The latent vectors, stored in edge storage, are sent to the central server at regular intervals or the storage space fills up to a certain level.
Then, at the central computer, they are restored to audio form by a decoder paired with the above encoder~\cite{azar_tradcomp_2019}.
At this time, even if the resolution of the collected sound sources is different, it is characterized by being restored to a similar Hi-Fi quality by the decoder of the central server.
The restored audio data is used for the purpose of developing a novel anomaly detection model or updating existing models.

To verify the proposed method, we collected data from three roads with different characteristics.
The audio is basically collected at 44,100 Hz.
For the purpose of saving storage space experiments in which downsampling is applied to assume a situation in which a microphone such as 11,025 Hz, and our zero-shot audio encoding method are also covered.

Overall, our contributions are summarized below:

\begin{itemize}
    \item When downsampling is applied, high-frequency information loss occurs, confirming that hinders precise anomaly detection. Our approach, zero-shot encoding and decoding minimize information loss and preserves anomaly detection performance at an appropriate level while maximizing storage efficiency.

    \item We show that there is no need to train new models to encode and decode for our domain, road noise. Our example deals with road anomaly detection as a target, but our method can also be extended to another edge computer-based data collection approaches in other domains.
\end{itemize}

\section{Related works}
\label{sec:relwork}

\subsection{Road condition identification}
\label{subsec:roadcond}
To identify abnormal situations on the road such as road bumps, cracks, or potholes, methods based on vision sensors have been proposed~\cite{ryu_camera_2015,rui_camera_2019, bibi_camera_2021,vojir_segmentation_2021}.
However, since the abnormal situation on the road is highly related to bad weather, and bad weather can obscure the view of the camera~\cite{ershadi_camcond_2017,qian_lidar_2021}, a vision sensor-based approach will show constrained detection performance.
Some approaches using motion or gyroscope sensors rather than a vision sensor can partially ease the above problem but it shows unstable performance that highly depends on pre-defined settings~\cite{salau_sensor_2018,sattar_sensor_2021}.
An audio-based anomaly detection method has been proposed as a way to overcome the problem of invisible situations to make decisions in bad weather or night situations~\cite{park_ncae_2022,park_foi_2023}.

To construct a reliable road anomaly detection system in outdoor environments, we inherit the above approach from prior research to take advantage of audio-based road anomaly detection.

\subsection{Data compression}
\label{subsec:datacomp}
The most intuitive way to maximize data storage efficiency is to collect data at a lower resolution.
However, when we need to create a high-quality anomaly detection model, we also need high-resolution data~\cite{zhao_dataquality_2021}. 

The computational approach which collects high-resolution data and encodes it into lower dimensions can be considered other than the above~\cite{azar_tradcomp_2019,lu_tradcomp_2021}.
When the encoding method is provided with a paired decoding method, we can easily compress the data into small sizes and decompress them into the original resolution.
In this case, some information loss may occur in the process of encoding and decoding, but it is recommended as an alternative to hardware-based capacity-saving methods that block information fundamentally.

An artificial neural network-based encoder-decoder (ED) shows better reconstruction performance than traditional methods~\cite{zhang_neuralcomp_2021,azar_neuralcomp_2022}.
In particular, a model trained for super-resolution (SR) purposes is useful as a method to help roughly estimate the high-frequency region of data collected at lower resolution~\cite{li_AST_unet_2022,han_NuWave_2022,alex_EnCodec_2022}.

We propose a storage space management method based on zero-shot encoding decoding using the pre-trained ED for SR purposes, considering that the resolution of audio sensors installed in each region may be different.

\section{Approach}
\label{sec:approach}

\begin{figure}[t]
    \resizebox{\columnwidth}{!}{%
        \hspace*{-0.45cm}
        \begin{tabular}{ccc}
            \includegraphics*[width=0.33\columnwidth,trim={0.0cm 0.0cm 0.0cm 0.0cm},clip]{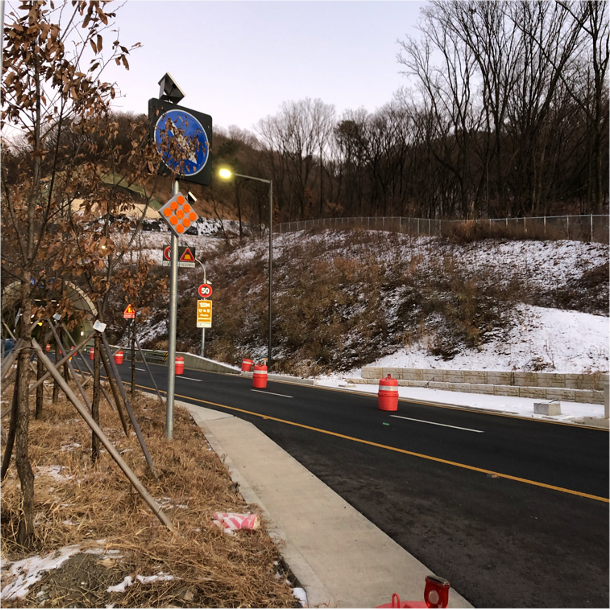} 
            & \includegraphics*[width=0.33\columnwidth,trim={0.0cm 0.0cm 0.0cm 0.0cm},clip]{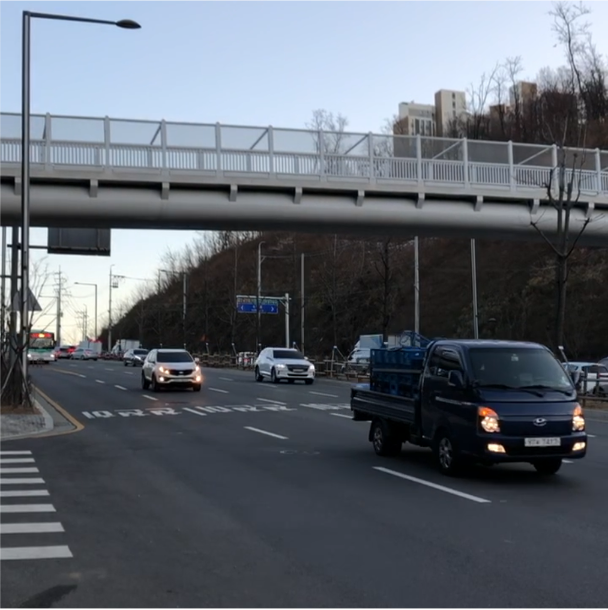}
            & \includegraphics*[width=0.33\columnwidth,trim={0.0cm 0.0cm 0.0cm 0.0cm},clip]{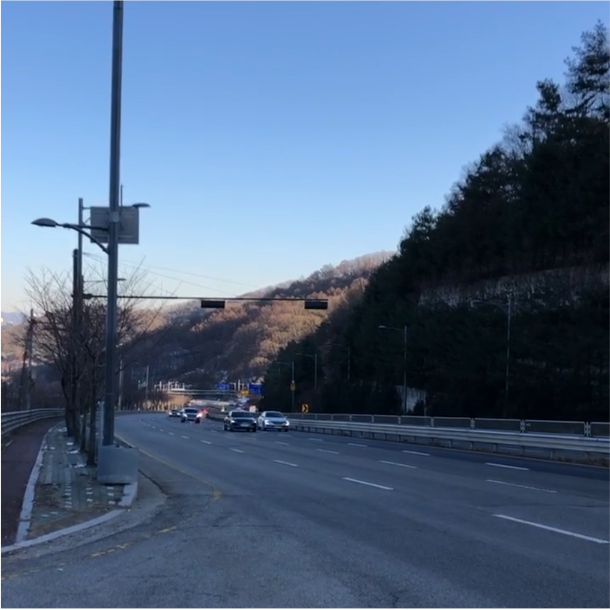} \\
            (a) Tunnel & (b) City & (c) Outer \\
        \end{tabular}
        \hspace*{-0.45cm}
    }
    \caption{Three data collection sites. Each has different road characteristics. The tunnel, shown in (a) has sound reverberation, and the city, (b), has irregular reflection by facilities. In the case of the outer road, (c), there is no sound reflection almost.}
    \label{fig:post}
\end{figure}

\subsection{Overview}
\label{subsec:overview}
The overall of our proposal is shown in Fig.~\ref{fig:scheme}. which is an encoding and decoding system for storing as much audio data as possible in an edge computer.
We separate pre-trained ED into each component encoder and decoder.
Then, we locate each of the above on the edge and central computers respectively.
Our approach only uses a single encoder and decoder pair rather than having each model for each different road environment as shown in Fig.~\ref{fig:post}.
This is dubbed as a zero-shot inference that can eliminate the hassle of preparing models to reflect the surrounding environment of countless sensors installed at outdoor points.

\subsection{Audio super-resolution}
\label{subsec:zshotenc}
For data compression, we utilize a pre-trained ASR model, EnCodec~\cite{alex_EnCodec_2022}, which is constructed with an encoder and decoder.
The sensors installed at each site to perform the road anomaly detection that we will cover may include expensive high-resolution microphones or low-resolution microphones, depending on the management budget of the local government.
The advantage of using the ASR model is that the input data can be converted into high-resolution data regardless of the input resolution.
This allows audio data collected from microphones of arbitrary resolutions as aforementioned can be integrated into Hi-Fi audio.

Any ASR model can be employed for the encoding and decoding process, but a structure in which the encoder and decoder can be used separately is recommended.
A method of performing information augmentation in a feature map or latent vector stage may rather increase the capacity of encoded data~\cite{li_AST_unet_2022}, so it should be avoided.

\subsection{Zero-shot encoding and decoding}
\label{subsec:zshotenc}
It is difficult to obtain a pre-trained ASR model on the road noise domain because the audio corresponding to the friction noise between the tire and the road surface, which we deal with, is not commonly used data.
In addition, the number of sensors installed on the roads we deal with is numerous and their characteristics are highly diverse, so it takes a huge amount of time and cost to build a model while guaranteeing the generalization ability.

As a way to easily overcome these methods, we adopt a zero-shot inference that utilizes a highly generalized ASR model which pre-learned with a wide range of audio data including general audio, speech, and music.
Following the above, we propose a method to separate the encoder of the pre-trained ASR model, place it on the edge computer and encode all the collected data.
The encoded data will be transmitted o the central computer and decoded.

\section{Experiments}
\label{sec:exps}

\subsection{Dataset}
\label{subsec:dataset}

To validate the zero-shot inference-based method, we should deal with varied data from different road environments.
Among the collectible road points, we selectively use three points with significantly different environmental characteristics as shown in Fig.~\ref{fig:post}.
We have collected the audio samples for four weather conditions at three locations as summarized in Table~\ref{tab:dataset}. 

\begin{table}[t]
    \centering
    \caption{Summary of the dataset acquired from the three different locations, in number of audio (number of driving events) form. Each audio is a 10-minute length and 10-second length driving events are extracted from the audio.}
    \begin{tabular}{l|rr|rrrrrr}
        \hline
            \multirow{2}{*}{\textbf{Post}} & \multicolumn{2}{c|}{\textbf{Normal}} & \multicolumn{6}{c}{\textbf{Abnormal}} \\
            \cline{2-9}
             & \multicolumn{2}{c|}{\textbf{Dry}} & \multicolumn{2}{c}{\textbf{Wet}} & \multicolumn{2}{c}{\textbf{Slush}} & \multicolumn{2}{c}{\textbf{Snow}} \\
        \hline
        \hline
            Tunnel & 10 & (384) & 10 & (21) & 4 & (7) & - & - \\
        \hline
            City & 10 & (804) & 10 & (529) & 2 & (11) & - & - \\
        \hline
            Outer & 10 & (1,153) & 9 & (1,032) & 10 & (76) & 3 & (5)\\
        \hline
        \hline
            Total & 30 & (2,341) & 29 & (1,582) & 16 & (94) & 3 & (5) \\
        \hline
    \end{tabular}
    \label{tab:dataset}
\end{table}


\subsection{Zero-shot compression}
\label{subsec:zerocomp}
\begin{figure}[t]
    \resizebox{\columnwidth}{!}{%
        \begin{tabular}{cc}
            & $X$ \\
            \rotatebox{90}{\qquad{}\quad{}\ $f_{44}$}  & \includegraphics*[width=1.0\columnwidth,trim={0.3cm 0.25cm 0.4cm 0.74cm},clip]{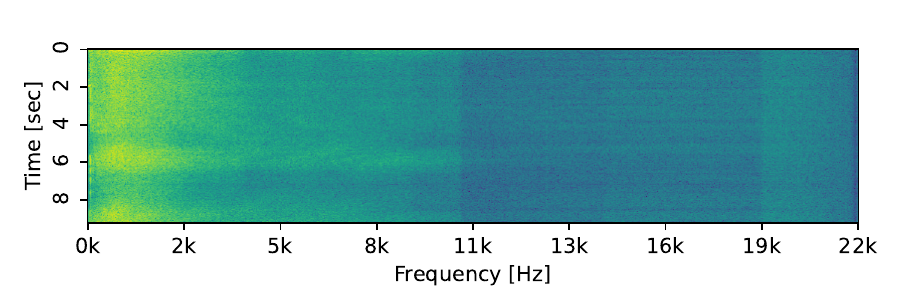} \\
            \rotatebox{90}{\qquad{}\quad{}\ $f_{22}$}  & \includegraphics*[width=1.0\columnwidth,trim={0.3cm 0.25cm 0.4cm 0.74cm},clip]{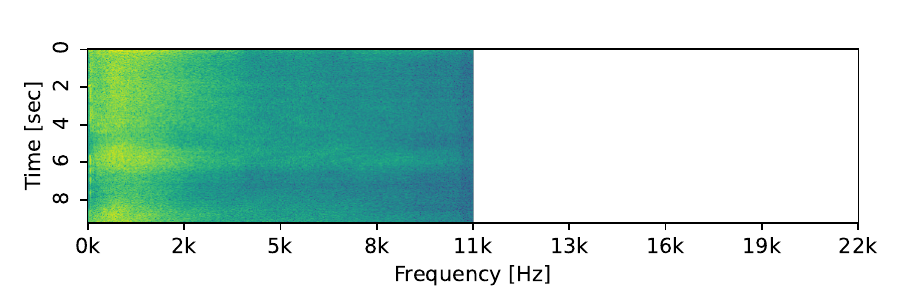} \\
            \rotatebox{90}{\qquad{}\quad{}\ $f_{11}$}  & \includegraphics*[width=1.0\columnwidth,trim={0.3cm 0.25cm 0.4cm 0.74cm},clip]{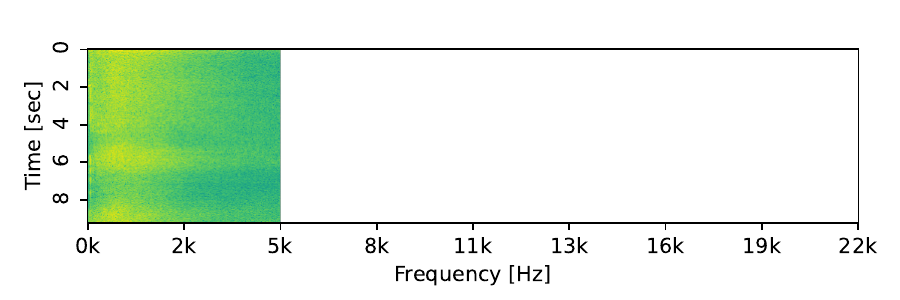} \\

            \noalign{\vskip 5mm} 

            & $D(E(X))$ \\
            \rotatebox{90}{\qquad{}\quad{}\ $\hat{f}_{44}$}  & \includegraphics*[width=1.0\columnwidth,trim={0.3cm 0.25cm 0.4cm 0.74cm},clip]{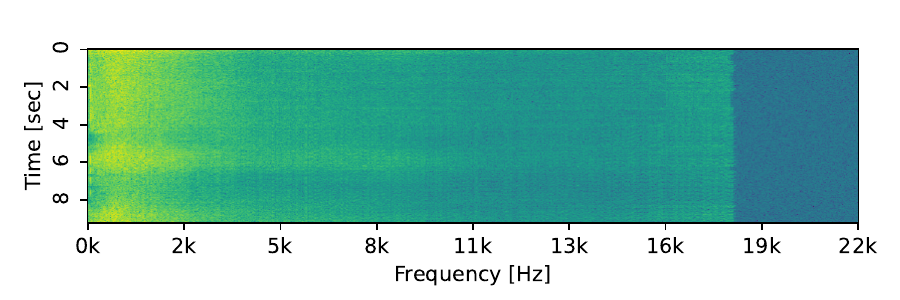} \\
            \rotatebox{90}{\qquad{}\quad{}\ $\hat{f}_{22}$}  & \includegraphics*[width=1.0\columnwidth,trim={0.3cm 0.25cm 0.4cm 0.74cm},clip]{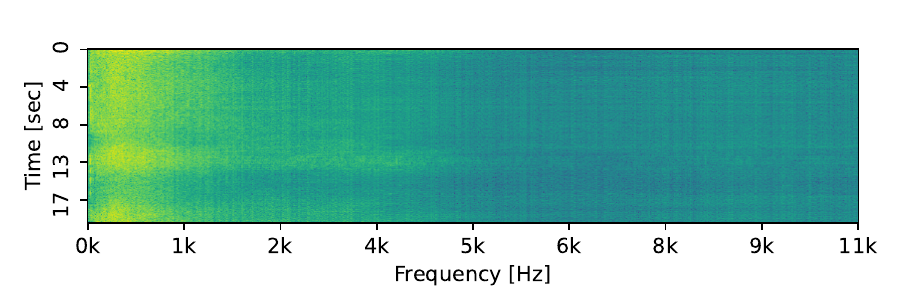} \\
            \rotatebox{90}{\qquad{}\quad{}\ $\hat{f}_{11}$}  & \includegraphics*[width=1.0\columnwidth,trim={0.3cm 0.25cm 0.4cm 0.74cm},clip]{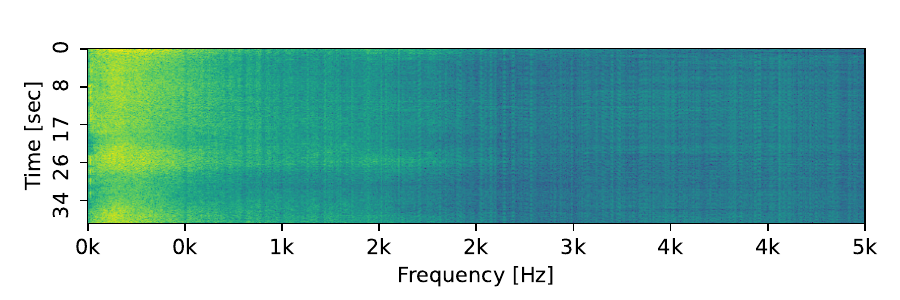} \\
            
        \end{tabular}
        \hspace*{-0.45cm}
    }
    \caption{Results of encoding and decoding for each audio input $X$ with EnCodec~\cite{alex_EnCodec_2022}. Each encoding and decoding is abbreviated as E and D.}
    \vspace*{-0.4cm}
    \label{fig:spectrogram}
\end{figure}

Referring to our purpose, maximizing data compression, it is important to minimize the restoration error as well as the compression capacity of the data.
Note that, we abbreviate 44,100 Hz, 22,050 Hz, and 11,025 Hz as
$f_{44}$, $f_{22}$, and $f_{11}$ respectively.
The $f_{22}$ and $f_{11}$ in Fig.~\ref{fig:spectrogram} show the fundamental loss of high-frequency components compared to $f_{44}$ which the sampling rate is set as low to reduce the file size.
Therefore, it should be avoided to set the low sampling rate with a hardware-based approach, and a method of collecting and compressing Hi-Fi data needs to be used in a computational approach.

When applying a pre-trained ASR model for audio compression and decompression, it shows not only preserving high-frequency components but also less information loss than the hardware-based approach as shown in $\hat{f}_{44}$ in Fig.~\ref{fig:spectrogram} and Table~\ref{tab:filesize}.

Through this experiment, we confirm that the data compression method based on the ASR model can increase the total amount of samples saved in an edge computer by $34.6\times$ while minimizing information loss. 
This means that the cost of data transmission can also be $34.6\times$ reduced.

\begin{table}[t]
    \centering
    \caption{Measured file size of one-second length audio data for each file management setting}
    \begin{tabular}{l||c||cc|c}
        \hline
            \textbf{Source} & \textbf{Hi-Fi} & \multicolumn{2}{c|}{\textbf{Lo-Fi}} & \textbf{ASR} \\ 
        \hline
            \textbf{Frequency} & \textbf{$f_{44}$~\cite{park_foi_2023}} & \textbf{$f_{22}$} & \textbf{$f_{11}$} & \textbf{$f_{44}$} \\
        \hline
        \hline
            File size & 173 KiB & 87 KiB & 44 KiB & 5 KiB \\
        \hline
            Ratio$_{size}$ $\downarrow$ & 1.000 & 0.503 & 0.254 & \textbf{0.029} \\
        \hline
    \end{tabular}
    \label{tab:filesize}
\end{table}

\subsection{Anomaly detection}
\label{subsec:anodet}

We simulate the compressed data collecting situation by the zero-shot encoding method in the central computer to check whether an anomaly detection model can perform at the appropriate level when it is trained with the decompressed dataset.
It is clear that zero-shot encoding is a method that can minimize information loss while maximizing compression rate compared to others, but considering that there is a slight difference from the original, we can estimate that anomaly detection performance may also be decreased.
Considering this, the method with the least performance degradation can be considered as the optimal method.

For the experiments, we downsample each audio by 2 and 4 scales to simulate the same Hi-Fi data as collected at Lo-Fi conditions.
Note that, the resolution of the original audio is 44,100 Hz, resolution for each downsampled audio is 22,050 Hz and 11,025 Hz respectively. 
Also, the compressed and decompressed audio data are used to verify the ASR case.
The measured anomaly detection performance with the area under the receiver operating characteristic curve (AUROC)~\cite{fawcett_auroc_2006} is summarized in Table~\ref{tab:performance_ad}.
In the case of using ASR, the average performance decreased to 92\%-level compared to the original Hi-Fi audio case.
However, we confirm that the anomaly detection performance of our method is more compliant than Lo-Fi.

\begin{table}[t]
    \centering
    \caption{Anomaly detection performance at each audio frequency.}
    \begin{tabular}{l||c||cc|c}
        \hline
            \textbf{Source} & \textbf{Hi-Fi} & \multicolumn{2}{c|}{\textbf{Lo-Fi}} & \textbf{ASR} \\ 
        \hline 
            \textbf{Frequency} & \textbf{$f_{44}$~\cite{park_foi_2023}} & \textbf{$f_{22}$} & \textbf{$f_{11}$} & \textbf{$f_{44}$} \\
        \hline
        \hline
            Tunnel & 0.963 & 0.961 & 0.957 & \textbf{0.967} \\
            City & 0.871 & 0.752 & 0.794 & \textbf{0.831} \\
            Outer & 1.000 & 0.841 & 0.818 & \textbf{0.847} \\
            Merge & 0.915 & 0.803 & 0.791 & \textbf{0.812} \\
        \hline
        \hline
            Average & 0.937 & 0.839 & 0.840 & \textbf{0.864} \\
        \hline
            Ratio$_{AUROC}$ $\uparrow$ & 1.000 & 0.895 & 0.896 & \textbf{0.922} \\
        \hline
    \end{tabular}
    \label{tab:performance_ad}
\end{table}

\subsection{Resolution integration}
\label{subsec:resint}
We verify that it can be integrated into equal-level of high-resolution audio through the encoding and decoding process when the resolution (sampling rate) of the collected audio is different.
If this premise is satisfied, it can guarantee that the proposed data compression and restoration framework via the ASR model works properly no matter which audio resolution.

If the data collected at low-resolution can be converted into high-resolution, it can be helpful to build a high-performance anomaly detection model considering the case where low-cost and low-resolution microphones are inevitably installed according to the budget.
When the three samples, assuming original high-resolution data and low-resolution data, are upsampled through the ASR model, they show almost the same difference from the original as summarized in Table~\ref{tab:recon_loss}.
Thus, we confirm that data of arbitrary resolution can be integrated into high-quality audio at the central server.

\begin{table}[t]
    \centering
    \caption{Reconstruction error, mean squared error (MSE), for each source audio frequency.}
    \begin{tabular}{l|ccc}
        \hline
            \textbf{Frequency} & \textbf{$f_{44}$} & \textbf{$f_{22}$} & \textbf{$f_{11}$} \\
        \hline
        \hline
            MSE$(f,\hat{f})$ & 71.07297 & 71.07244 & 71.07209 \\
        \hline
    \end{tabular}
    \label{tab:recon_loss}
\end{table}

\section{Conclusion}
\label{sec:conclusion}
We propose a method based on the pre-trained ASR model for storing as many audio samples as possible in an edge computer with limited storage capacity installed for the purpose of road anomaly detection.
Our method shows that adequate performance can be obtained by using only one generalized encoder-decoder pair instead of each encoder-decoder corresponding to each post or type of road with high environmental diversity.
Each audio is highly compressed from its original size of 173 KiB per second to 5 KiB, showing that it can store up to $34.6\times$ as many audio samples.
In addition, even if an anomaly detection model is trained by collecting compressed audio samples at the central computer, an appropriate level can be achieved.
Some degradation of the anomaly detection performance is caused by a slight information loss during the encoding and decoding process but there is room for improvement via proper encoder-decoder pairs.
In future work, we plan to explore the encoder-decoder model with better generalization ability or trained on the road noise domain as a way to construct more stable systems.

\section*{Acknowledgements}
We are grateful to all the members of SK Planet Co., Ltd., who have supported this research, providing equipment for the experiment.

\balance

\end{document}